\documentclass[twocolumn,aps,pra,superscriptaddress,tightenlines]{revtex4}
\usepackage{amsmath}
\usepackage{amsfonts}
\usepackage{graphicx}
\usepackage{epsfig}
\usepackage{color}
\usepackage[colorlinks,citecolor=blue]{hyperref}

\begin{document}

\title{Steady-state subradiance manipulated  by the two-atom decay}
\author{Qian Bin}
\affiliation{School of Physics and Institute for Quantum Science and Engineering, Huazhong University of Science and Technology, Wuhan, 430074, China}


\author{Xin-You L\"{u}}
\email{xinyoulu@hust.edu.cn}
\affiliation{School of Physics and Institute for Quantum Science and Engineering, Huazhong University of Science and Technology, Wuhan, 430074, China}

\date{\today}

\begin{abstract}
We investigate theoretically the collective radiance characteristics of an atomic ensemble with the simultaneous decay of two atoms. We show that the two-atom decay can significantly  suppress the steady-state collective radiance of the atoms, expanding the region of subradiance.  In the steady-state subradiance regime,  the system is in an entangled state, and the mean populations of the system in the excited state and ground state of the atoms  are almost equal. The processes of the two-atom decay can be demonstrated by the population distribution of the system state on the Dicke ladder.  Moreover, we show the  correlation property of the emitted light from the atomic ensemble, where the correlation function is rewritten in the presence of the two-atom decay. We find that  the emitted photons of  steady state only show bunching in the case of two-atom decay. This work broadens the realm of collective radiance, with potential applications for quantum information processing. 
 \end{abstract}


\maketitle

\section{introduction}
Collective spontaneous emission is one of the central topics of modern optics. An intriguing example exhibiting collective effect is superradiance discorvered by Dicke in 1954, where radiance intensity from an ensemble of emitters is enhanced due to the constructive interference between the radiances from individual emitters\,\cite{Dicke1954}. Superradiance behavior was first observed experimentally more than four decades ago\,\cite{Skribanowitz1973HM,Friedberg1973HT}. To date, superradiance has become a useful resource in lasing engineering\,\cite{Chan2003BV,Norcia2016WC,Bohnet2012CW}, precision measurements\,\cite{Kim2006BO, Rohlsberger2010SS, Norcia2018CM}, quantum memories\,\cite{Walther2019SK}, and quantum information\,\cite{Kuzmich2003BB,Casabone2015FB}.  Its counterpart, subradiance, describes the cooperative suppression of  spontaneous emission from an ensemble of emitters\,\cite{Dicke1954}. Compared with superradiance, it is significantly hard to experimentally observe subradiance effect, as the subradiance states are weakly coupled to the radiative vacuum and are rather sensitive to non-radiative decoherence. Direct observations of subradiance have been achieved  in a pair of ions\,\cite{DeVoe1996Brewer} and molecular systems\,\cite{Hettich2002SZ,Takasu2012ST,McGuyer2015MI}. Recently, subradiance was also observed in cold atomic clouds\,\cite{Guerin2016AK,Cipris2021ME,Das2020LY, Gold2022HY,Ferioli2021GH}, superconducting circuit systems\,\cite{Wang2020LF}, Rydberg atoms\,\cite{Stiesdal2020BK}, and 2D layer of atoms\,\cite{Rui2020WRA}. Because of its special radiation characteristics, subradiance may have important applications in quantum metrology\,\cite{Ostermann2013RG} and quantum information processing\,\cite{Asenjo-Garcia2017MCA,Facchinetti2016JR,Needham2019,Guimond2019GV}. For example, subradiance can be used to prolong the stored lifetimes of information by its slow collective emission. 

Usually superradiance and subradianc are studied in the pulsed regime, where the emitters initially prepared in the excited state rapidly relax to the ground states with the single-atom decay channel and then the radiance terminates\,\cite{Gross1982Haroche}. This collective radiance  is transient.  It has recently been proposed that the superradiance and subradiance can be obtained in steady state,  where both continuous dissipation and pumping have been considered in the systems\,\cite{Bohnet2012CW,Meiser2010Holland1, Meiser2010Holland2, Gegg2018CK, Shankar2021RJ, Auffeves2011GP, Dorofeenko2013ZV, Jager2021LS, Zheng2016Cooper, Zhang2018ZM, Kirton2017Keeling, Patra2019AY, Xu2016JS,Bohnet2014CW}. The emitters collectively emit photons and can be repumped to provide a steady supply for the system\,\cite{Bohnet2012CW,Meiser2010Holland1, Meiser2010Holland2,Shankar2021RJ}. This system can continuously generate collective light emission with the single-atom decay channel. However, the present studies on superradiance and subradiance of steady state are confined to considering only the single-atom decay as the collective decay of the atomic ensemble. Recently, it has been proposed that collective light emission with the single-atom decay channel is  suppressed and the two-atom decay channel (i.e., the simultaneous decay of two atoms of an ensemble) emerges in waveguide quantum electrodynamics (QED) when the emitter frequencies are below the edge of the propagation bound\,\cite{Wang2020JK}. Reference\,\cite{Qin2021MJ} proposed that the quantum degenerate  parametric amplification in a cavity QED system can also cause the two-atom decay.  The two-atom decay can lead to supercorrelated radiance with perfectly correlated spontaneous emission \,\cite{Wang2020JK} and the generation of a long-lived macroscopically quantum superposition state\,\cite{Qin2021MJ}.  Then one question that arises naturally is whether the two-atom decay could influence the collective radiance characteristics of steady state of an atomic ensemble. 

Here, we study the steady-state collective radiance of an incoherently pumped atomic ensemble.  The atomic ensemble can undergo two-atom collective decay via the cavity. We find that, compared with the case of single-atom decay,  two-atom decay can significantly suppress the steady-state collective radiance of the atomic ensemble, expanding the region of subradiance.  In the subradiance regime  manipulated by the collective decay, the system is in an entangled state, and the mean populations of the system in the excited state and ground state of the atoms  are almost equal. We also show the state spaces in different radiance regimes, which clearly demonstrate the processes of the two-atom decay and the  collective radiance characteristics of the atoms manipulated  by the two-atom decay. Moreover, we investigate the correlation characteristics of the emitted photons from the atomic ensemble. Nearly coherent emitted photons can be obtained in the superradiance regime when only single-atom decay is considered, but does not occur in the case of including only two-atom decay. In the latter case, the emitted photons of steady state only show bunching in the subradiance, superradiance, and uncorrelated radiance regimes, where the correlation function is rewritten due to the cavity field $\hat{a}\propto\hat{J}_-^2$.   

Compared to earlier works on the application of two-atom decay\,\cite{Wang2020JK,Qin2021MJ}, here we investigate a different quantum effect, i.e., steady-state subradiance of an atomic ensemble manipulated by two-atom decay. The subradiance in steady state originates from the competition between the collective decay and the repumping on the atoms, where the system makes the balance between the collective decay and the weak repumping on atoms  by a suppressed emission of the atomic ensemble. This subradiance may be used for quantum storage due to its slow collective emission. In contrast, Ref.\,\cite{Wang2020JK} studied supercorrelated radiance, where the two-atom decay makes perfectly correlated  spontaneous emission   and can lead to collective acceleration beyond the $N^2$ scaling of superradiance;  Ref.\,\cite{Qin2021MJ} studied the generation of a long-lived macroscopically quantum superposition state  by the two-atom decay. They may have potential applications in lasing engineering and noise-immune quantum technologies. The collective radiance of steady state means that the system stably emits photons through continuous dissipation and pumping, which is also essentially different from supercorrelated radiance, where the emitters initially prepared in the excited state rapidly relax to the ground states and then the radiance terminates.  Our work not only expands the realm of the two-atom decay by bringing it to the next stage of application in steady-state collective radiance,  but also is fundamentally interested in exploring collective radiance theory.

\begin{figure}
\includegraphics[width=8.6cm]{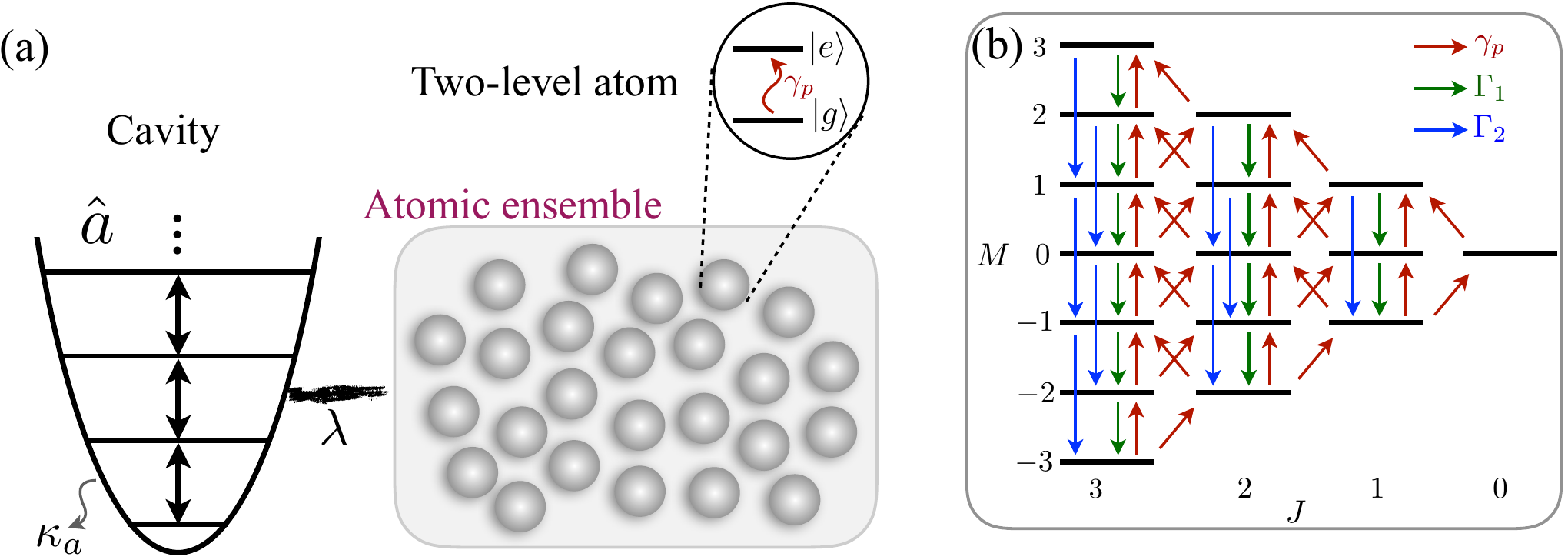}\\
\caption{(a) Schematic of the model. The cavity field is coupled to the atomic ensemble consisting of $N$ identical two-level atoms, where $\lambda$ is the collective coupling strength between the cavity and atomic ensemble, and $\kappa_a$ is the rate of cavity field. The atoms are incoherently repumped with pump rate  and can undergo the two-atom collective decay via the cavity field. (b) Dicke space for $N=6$ showing the Dicke states $|J,M\rangle$. The red, green, and blue arrows correspond to the processes of repuming $\gamma_p$, single-atom decay $\Gamma_1$, and two-atom decay $\Gamma_2$, respectively. }\label{fig1}
\end{figure}

\section{Model}
We consider an atomic ensemble that consists of $N$ identical two-level atoms, each with an excited state $|e\rangle$ and  ground state $|g\rangle$.  
As shown in Fig.\,\ref{fig1}(a), it is considered that all  atoms are collectively coupled to a cavity field and the coupled atom-cavity system is described by  the Hamiltonian $H=\lambda(\hat{a}^\dag \hat{J}_-^2+\hat{a}\hat{J}_+^2 )$, where  $\hbar=1$, $\hat{J}_-=\sum_{n=1}^N\hat{\sigma}_n=(\hat{J}_+)^\dag$ is the collective lowering operator,  $\hat{\sigma}_n^\dag=|e\rangle\langle g|$ is the Pauli creation operator for the $n$th two-level atom, and $\lambda$  is the effective atom-cavity collective coupling strength. Here, we have considered $\omega_a\approx2\omega_{\sigma}$, and $\omega_a$  and $\omega_{\sigma}$ are the frequencies of the cavity and two-level atom, respectively. The model can be achieved  in a nonlinear cavity QED system with degenerate parametric amplification\,\cite{Qin2021MJ}. The dissipative dynamics of the  coupled system can be implemented with a Lindblad-type master equation 
\begin{align}\label{eq001}
\frac{d \hat{\rho}}{dt}=-i[\hat{H},\hat{\rho}]+\gamma_p \sum_{n=1}^N \mathcal{L}[\hat{\sigma}_n^\dag]+ \kappa_a \mathcal{L}[\hat{a}],
 \end{align}
where the Liouvillian superoperator $\mathcal{L}$ is defined as $\mathcal{L}[\mathcal{\hat{O}}]=(2\mathcal{\hat{O}}\hat{\rho}\mathcal{\hat{O}}^\dag-\hat{\rho}\mathcal{\hat{O}}^\dag\mathcal{\hat{O}}-\mathcal{\hat{O}}^\dag\hat{\mathcal{O}}\hat{\rho})/2$, and $\kappa_a$ and $\gamma_p$ represent the  decay rate of the cavity and the rate of incoherent pumping of the individual atoms, respectively.  The dissipation of the system is balanced by pumping the atoms to the excited states from their ground states. This pumping on the atoms can be regarded as spontaneous absorption from  $|g\rangle$ to $|e\rangle$. It can be achieved experimentally  by optically driving a Raman transition from the ground state $|g\rangle$ to an auxiliary excited state that can rapidly decay to the excited state $|e\rangle$\,\cite{Bohnet2012CW,Meiser2010Holland1}. The spontaneous emission of the individual atoms has been neglected here, since the decay rate of the independent atoms $\gamma$ is much less than that of the cavity, i.e., $\kappa_a\gg\gamma$.  In  this limit of  bad-cavity $\kappa_a\gg\gamma$, the cavity mode $\hat{a}$ can be adiabatically eliminated\,\cite{Meiser2010Holland2}. The emission of the cavity photons is thus characterized by the collective emission of the atomic ensemble, with $\hat{a}\propto\hat{J}_-^2$ in the system.   The above Eq.\,(\ref{eq001}) can be reduced to an effective master equation of collective radiance 
\begin{align}\label{eq002}
\frac{d \hat{\rho}}{dt}=\gamma_p \sum_{n=1}^N \mathcal{L}[\hat{\sigma}_n^\dag]+ \Gamma_2 \mathcal{L}[\hat{J}_-^2], 
 \end{align}
where the two-atom decay emerges in the system due to  $\hat{a}\propto\hat{J}_-^2$, and the collective decay rate of the atoms $\Gamma_2=4\lambda^2/\kappa_a$. As a comparison, we also consider the system containing only single-atom decay, here the last term of Eq.\,(\ref{eq002}) is replied by $ \Gamma_1 \mathcal{L}[\hat{J}_-]$, and $\Gamma_1$ is the rate of the single-atom decay.  This case can be easily achieved in a Tavis-Cumming model under the limit of bad cavity\,\cite{Bohnet2012CW}. 

To understand the behavior of the collective emission of the atomic ensemble more conveniently, we discuss the dynamics of the system in the collective basis $|J,M\rangle$, with the quantum numbers $J=0,1,2,\dots,N/2$ (for an even number $N$) and $M=-J,-J+1,\dots,J-1,J$. The state $|J,M\rangle$ is the joint eigenstate of the operators $\hat{\bold{J}}^2$ and $\hat{J}_z$, with $\hat{\bold{J}}^2|J,M\rangle=J(J+1)|J,M\rangle$ and $\hat{J}_z|J,M\rangle=M|J,M\rangle$, where $\hat{J}_z=\sum_{n=1}^N \hat{\sigma}_n^j/2$ ($j=x,y,z$), $\hat{\sigma}_n^x=\hat{\sigma}_n^\dag + \hat{\sigma}_n$,  $\hat{\sigma}_n^y=i(\hat{\sigma}_n - \hat{\sigma}_n^\dag)$, and  $\hat{\sigma}_n^z=\hat{\sigma}_n^\dag\hat{\sigma}_n - \hat{\sigma}_n\hat{\sigma}_n^\dag $. The value of small $J$ corresponds to subradiant subspace. The discrete Dicke state space for the collective atomic states is shown in Fig.\,\ref{fig1}(b). In this state space, the single-atom and two-atom collective decays give rise to the transitions with the differences of the quantum number $\delta M = M-M'=-1$ (see the green arrows) and $\delta M=-2$  (see the blue arrows) within a ladder of a constant $J$, respectively. The pumping on individual atoms can generate the transition with $\delta M=1$ within the ladder of $J$ and its adjacent ladder of $J\pm1$, corresponding to the red arrows in Fig.\,\ref{fig1}(b).

\section{Results}
To investigate in detail the influence of the two-atom decay on the steady-state collective radiance of the atomic ensemble, we calculate  the average occupation of the emitters and atom-atom correlation $R_f$ in the two cases of  single-atom decay and two-atom decay. The atom-atom correlation is given by 
\begin{align}\label{eq003}
R_f=\frac{1}{N}\langle \hat{J}_+ \hat{J}_-\rangle- \frac{1}{N}\sum_{n=1}^N\langle \hat{\sigma}_n^\dag \hat{\sigma}_n\rangle,
 \end{align}
where $\langle \hat{J}_+ \hat{J}_-\rangle$  and $\sum_{n=1}^N\langle \hat{\sigma}_n^\dag \hat{\sigma}_n\rangle$ describe average collective occupation of the atoms and the total population from $N$ individual atoms, respectively. The effect of atom-atom correlations has been included in the collective occupation term $\langle \hat{J}_+ \hat{J}_-\rangle$.  Under this definition, $R_f=0$ indicates an uncorrelated feature between the atoms, where the collective occupation of the atomic ensemble is the sum of the populations of $N$ individual atoms, i.e.,  $\langle \hat{J}_+ \hat{J}_-\rangle=\sum_{n=1}^N\langle \hat{\sigma}_n^\dag \hat{\sigma}_n\rangle$.  $R_f>0$, i.e., $\langle \hat{J}_+ \hat{J}_-\rangle>\sum_{n=1}^N\langle \hat{\sigma}_n^\dag \hat{\sigma}_n\rangle$, means that the atom-atom correlation increases the collective population of the atoms. $R_f<0$, i.e., $\langle \hat{J}_+ \hat{J}_-\rangle<\sum_{n=1}^N\langle \hat{\sigma}_n^\dag \hat{\sigma}_n\rangle$, corresponds to the suppression of the collective population of the atoms by atom-atom correlation.

\begin{figure}
\includegraphics[width=8.6cm]{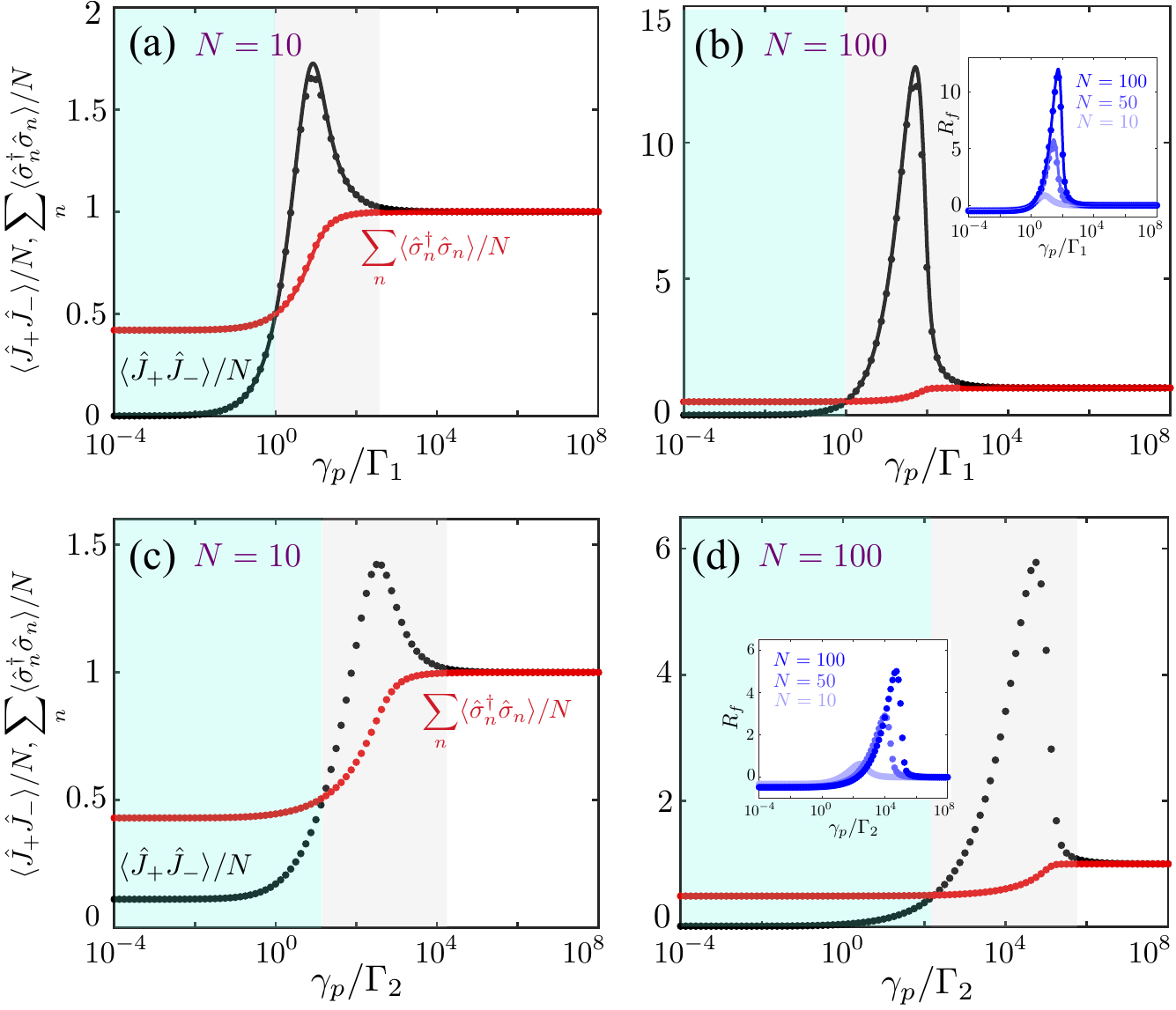}\\
\caption{Averaged population of the atoms in steady state vs $\gamma_p/\Gamma_1$ and $\gamma_p/\Gamma_2$ in the cases of including only the single-atom decay (a and b) and two-atom decay  (c and d), respectively. The blue and gray shaded areas indicate the increase and suppression of the population of the atoms by the atom-atom correlation, respectively, and the white areas indicate an uncorrelated feature between the atoms. Insets: atom-atom correlation $R_f$ vs $\gamma_p/\epsilon$ for different $N$. The solid lines and dots correspond to the analytical and numerical results, respectively. System parameters are (a, c) $N=10$ and (b, d) $N=100$.}\label{fig2}
\end{figure}

We first consider the case of single-atom decay,  where the equations of motion from the master equation Eq.\,(\ref{eq002}) are given by 
\begin{subequations}
\begin{align}\label{eq004}
&\frac{d}{dt}\!\langle \hat{\sigma}_1^z\rangle \!=\!-(\!\gamma_p\!+\!\Gamma_1\!)\!\langle \hat{\sigma}_1^z\rangle\!\!-\!2(\!N\!\!-\!1\!)\Gamma_1\!\langle\hat{\sigma}_1^\dag\hat{\sigma}_2\rangle\!+\!(\!\gamma_p\!-\!\Gamma_1\!),\\
&\frac{d}{dt}\langle \hat{\sigma}_1^\dag \hat{\sigma}_2\rangle=-(\gamma_p+\Gamma_1)\langle\hat{\sigma}_1^\dag\hat{\sigma}_2\rangle+\frac{\Gamma_1}{2}\langle \hat{\sigma}_1^z\hat{\sigma}_2^z\rangle+\frac{\Gamma_1}{2}\langle \hat{\sigma}_1^z\rangle\nonumber\\
&~~~~~~~~~~~~~~~+\Gamma_1(N-2)\langle \hat{\sigma}_1^z\hat{\sigma}_2\hat{\sigma}_3^\dag\rangle,\\
&\frac{d}{dt}\langle \hat{\sigma}_1^z\hat{\sigma}_2^z\rangle=-2(\gamma_p+\Gamma_1)\langle \hat{\sigma}_1^z\hat{\sigma}_2^z\rangle+2(\gamma_p-\Gamma_1)\langle \hat{\sigma}_1^z\rangle\nonumber\\
&~~~~~~~~~~~~~~~+4\Gamma_1\langle \hat{\sigma}_1^\dag \hat{\sigma}_2 \rangle-4\Gamma_1 (N-2)\langle\hat{\sigma}_1^z\hat{\sigma_2}\hat{\sigma}_3^\dag\rangle.
\end{align}
\end{subequations} 
Here we have considered $\langle \hat{\sigma}_n^\dag\hat{\sigma}_{n'}\rangle=\langle \hat{\sigma}_1^\dag\hat{\sigma}_{2}\rangle$ for all $n\neq n'$ due to the symmetry of the expectation values related to particle exchange\,\cite{Meiser2010Holland1, Meiser2010Holland2,Shankar2021RJ,Xu2016JS}. The above Eqs.\,(4a)-(4c) can be reduced to a closed set of equations by factorizing the third-order expectation values as  $\langle\hat{\sigma}_1^z\hat{\sigma_2}\hat{\sigma}_3^\dag\rangle \approx \langle\hat{\sigma}_1^z\rangle \langle \hat{\sigma_1}^\dag\hat{\sigma}_2\rangle$\,\cite{Meiser2010Holland1, Meiser2010Holland2,Shankar2021RJ,Xu2016JS}. This factorization might cause partial decorrelation between atoms, but complete decorrelation cannot occur, since the term $ \langle \hat{\sigma_1}^\dag\hat{\sigma}_2\rangle$ includes the effect of atom-atom correlation. In the  steady-state limit, we obtain
\begin{subequations}
\begin{align}\label{eq005}
&\langle \hat{\sigma}_1^\dag\hat{\sigma}_2\rangle=-\frac{c_2}{2c_1}+\frac{\sqrt{c_2^2-4c_1c_3}}{2c_1},\\
&\langle \hat{\sigma}_1^z\rangle=\frac{\gamma_p-\Gamma_1}{\gamma_p+\Gamma_1}+\frac{\Gamma_1(N-1) (c_2-\sqrt{c_2^2-4c_1c_3})}{c_1(\gamma_p+\Gamma_1)},
\end{align}
\end{subequations} 
where 
\begin{subequations}
\begin{align}\label{eq006}
&c_1=\frac{4(N-1)(N-2)\Gamma_1^2}{(\gamma_p+\Gamma_1)^2},\\
    &c_2=2+\frac{2N\Gamma_1}{\gamma_p+\Gamma_1}-\frac{2\Gamma_1(2N-3)(\gamma_p-\Gamma_1)}{(\gamma_p+\Gamma_1)^2},\\
&c_3=\frac{2\Gamma_1(\Gamma_1-\gamma_p)}{(\gamma_p+\Gamma_1)^2}.
\end{align}
\end{subequations} 
Then the solutions of the steady-state light emissions can thus be given by inserting Eqs.\,(5a)-(5b) into $\langle \hat{J}_+\hat{J}_-\rangle=N(\langle \hat{\sigma}_1^z\rangle-1)/2+N(N-1)\langle \hat{\sigma}_1^\dag\hat{\sigma}_2\rangle$ and  $\sum_{n=1}^N\langle \hat{\sigma}_n^\dag \hat{\sigma}_n\rangle=N(\langle \hat{\sigma}_1^z\rangle+1)/2$. Figures \ref{fig2}(a)-\ref{fig2}(b) show the population of the atoms and atom-atom correlation in the ensemble, obtained by the analytical solutions and numerically calculating  the master equation Eq.\,(\ref{eq002}). Here, Eq.\,(\ref{eq002}) can be directly calculated by the permutational-invariant quantum  solver\,\cite{Shammah2018AL} in QUTIP\,\cite{Johansson2012NN}.  The very good agreement between analytical solutions and fully numerical simulations demonstrates the validity of the above approximation.

\begin{figure}
\includegraphics[width=8cm]{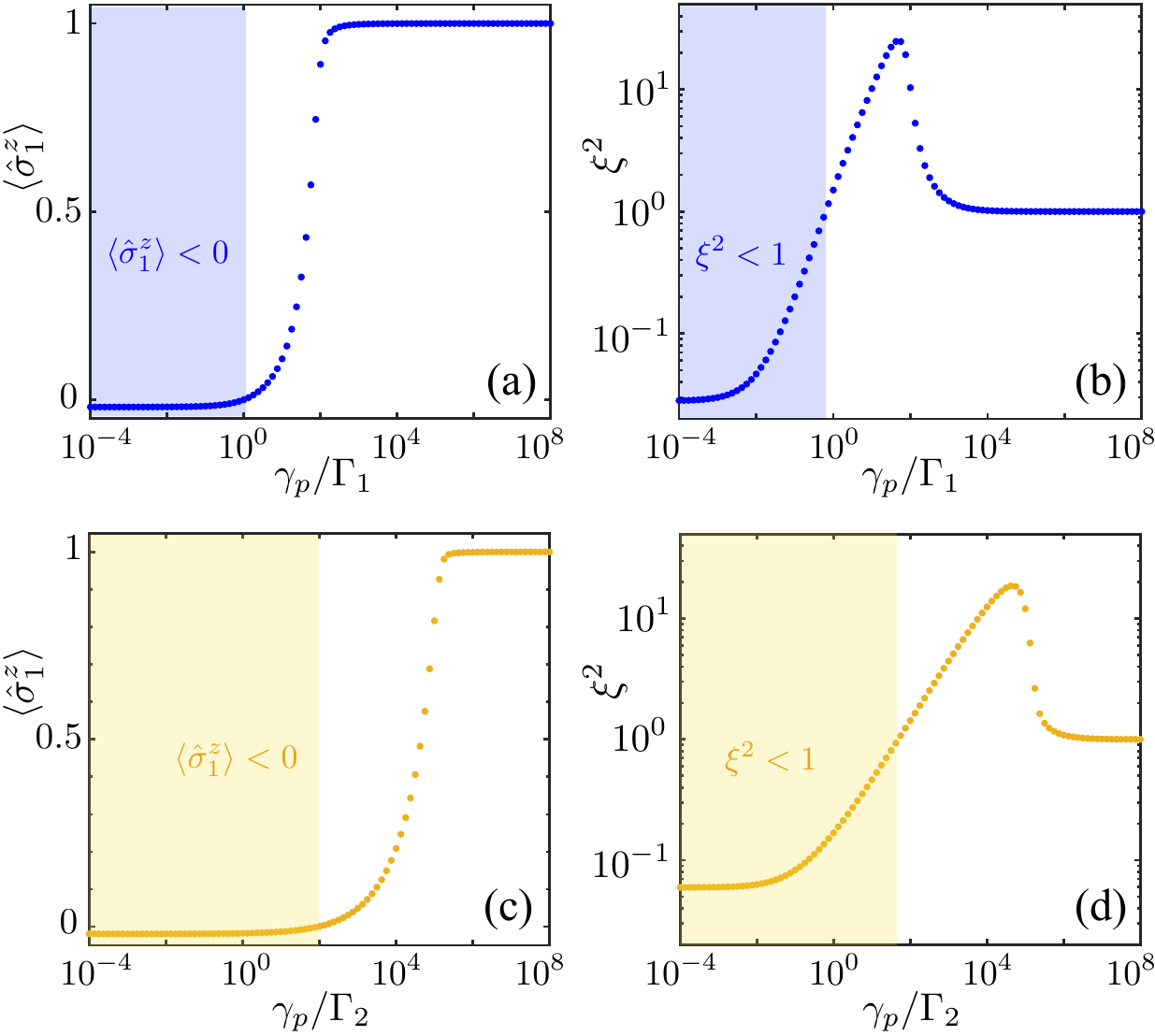}\\
\caption{ Mean atomic inversion $\langle \hat{\sigma}_1^z\rangle$ and squeezing parameter $\xi^2$ vs (a and b) $\gamma_p/\Gamma_1$ and (c and d) $\gamma_p/\Gamma_2$ when $N=100$. The blue and yellow dots correspond to the cases of single-atom decay and two-atom decay, respectively.}\label{fig3}
\end{figure}

 %
%
%
Now, let us consider the case of two-atom decay. In Figs.\,\ref{fig2}(c)-\ref{fig2}(d), we present the population of the atoms and atom-atom correlation in the ensemble, obtained by numerically solving the master equation Eq.\,(\ref{eq002}). 
Comparing Figs.\,\ref{fig2}(a)-\ref{fig2}(b) and \ref{fig2}(c)-\ref{fig2}(d),  shows that the two-atom decay can significantly suppress the collective population of the atoms in steady state in a wider parameter regime, leading to the expanding of the subradiance region.  This is because the system with two-atom decay relaxes to lower energy states faster than the one with single-atom decay.  More energy is needed to repump the atoms to their excited states  in the system with  two-atom decay,  compared with that in the case of single-atom decay, as shown in Figs.\,\ref{fig3}(a) and \ref{fig3}(c).  In the subradiance regime manipulated by the collective decay, i.e., single-atom decay or two-atom decay,  the system is in an entangled state, and  the mean populations of the system in the excited state and ground state of the atoms  are almost equal, as shown in Fig.\,\ref{fig3}. Here the entanglement can be adjusted by the witness
\begin{align}\label{eq007}
&\xi^2=\frac{2[(\Delta\hat{J}_x)^2+(\Delta\hat{J}_y)^2+(\Delta\hat{J}z)^2]}{N}, 
\end{align}
with $(\Delta\hat{J}_j)^2=\langle \hat{J}_j^2\rangle-\langle\hat{J}_j\rangle^2$ ($j=x,y,z$), and the spin squeezing parameter $\xi^2<1$ indicates the entanglement establishing.

\begin{figure}
\includegraphics[width=8.6cm]{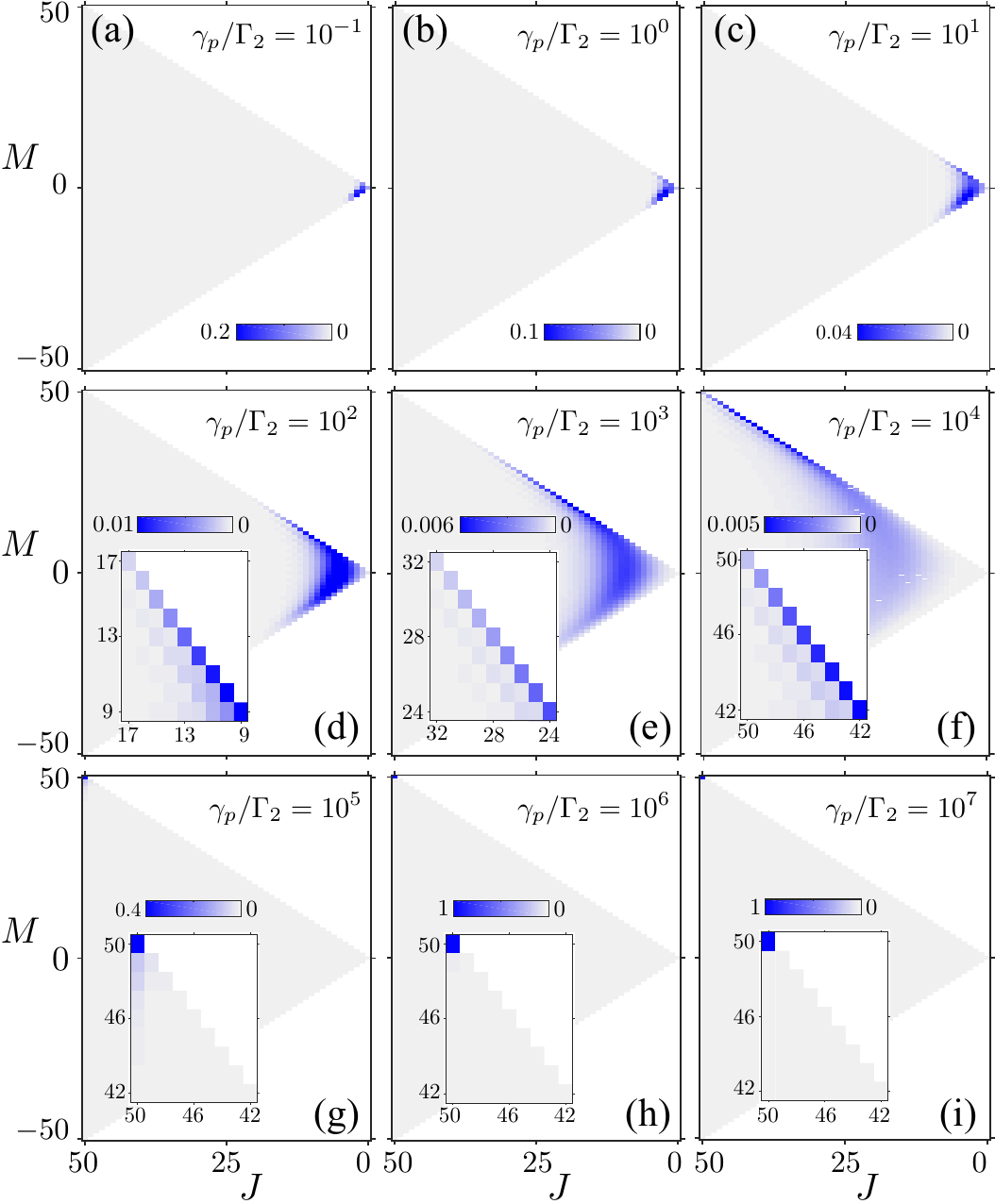}\\
\caption{Population distribution of system states on the Dicke ladder for different $\gamma_p/\Gamma_2$ when $N=100$: steady states of (a-c) subradiance, (d-g) supperradiance, and (h,i) uncorrelated radiance. Insets: Enlarged region of the population distribution.}\label{fig4}
\end{figure}

\begin{figure}
\includegraphics[width=8.6cm]{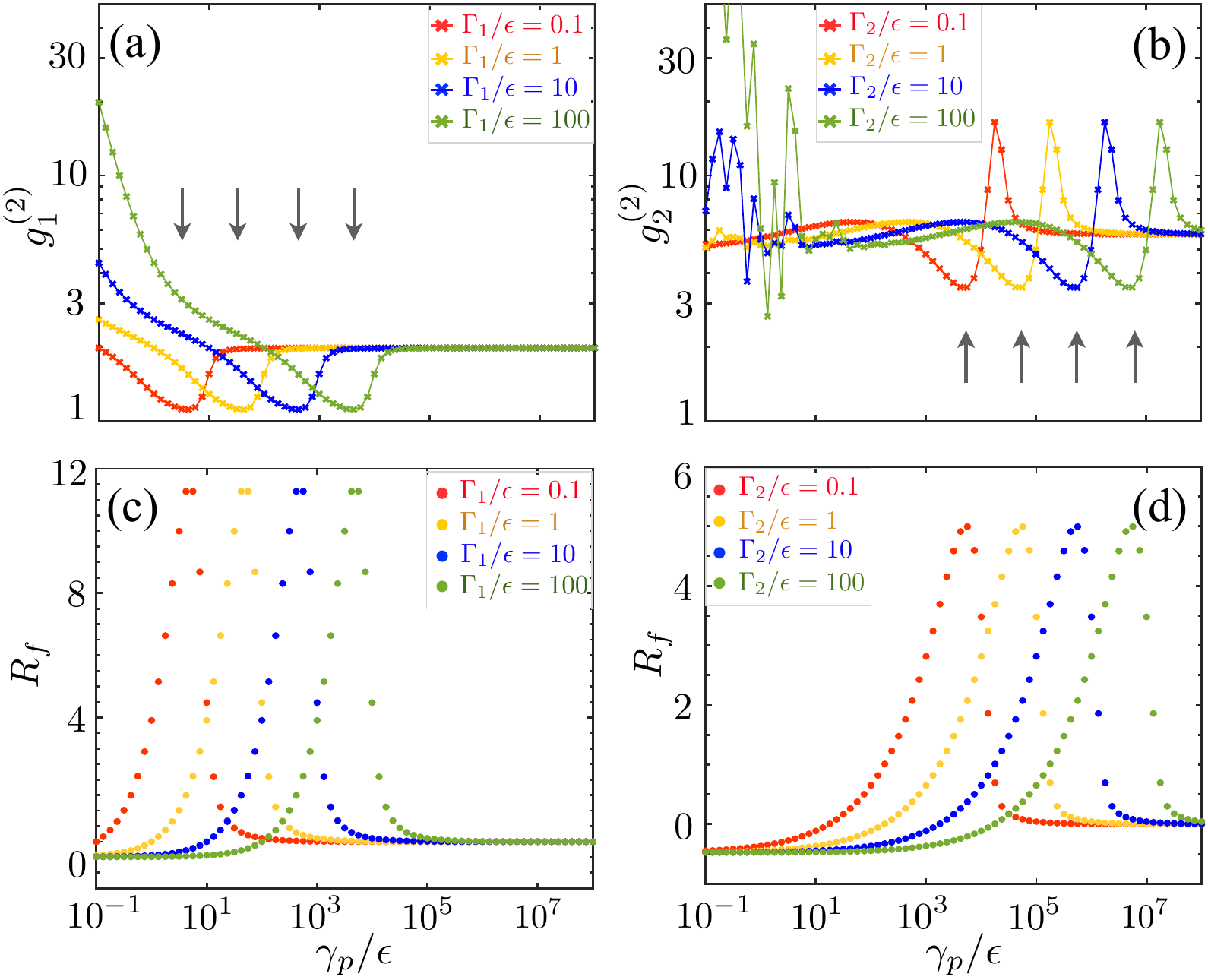}\\
\caption{(a and b) Equal-time second-order correlation function $g^{(2)}_i(0)$ ($i=1,2$) and (c and d) atom-atom correlation $R_f$ vs $\gamma_p/\epsilon$ for different $\Gamma_i/\epsilon$.  Panels (a,c) and (b,d) correspond to the cases of single-atom decay and two-atom decay, respectively. Other system parameters are $N=100$ and $\epsilon/2\pi=10\,{\rm Hz}$. }\label{fig5}
\end{figure}

In Figs.\,\ref{fig4}(a)-\ref{fig4}(i), we show the population distribution of the system state on the Dicke ladder in the case of two-atom decay. In the subradiance regime [see Figs.\,\ref{fig5}(a)-\ref{fig5}(c)],  the system evolves in the states with smaller and smaller $J$, since the pumping mainly drives the adjacent ladder transitions $|J\rangle\to|J-1\rangle$ when $M<0$. The populations of the system in the excited state and ground state of the individual atom in this steady state  are almost equal [also see Fig.\,\ref{fig3}(c)],   and the collective population of the atoms  is suppressed by atom-atom correlation. This steady state generates a suppressed emission. In the superradiance regime [see Figs.\,\ref{fig5}(d)-\ref{fig5}(g)], the collective two-atom decay is dominated in the system for the states with large $J$, while the repumping is dominated for the states with small $J$.  In this steady state, the population of the system in the excited state  of the atoms  is greater than that in the ground state [also see Fig.\,\ref{fig3}(c)], and the collective population of the atoms  is increased by atom-atom correlation.  The population distribution of the system states on the Dicke ladder for large $J$ also demonstrates that the two-atom decay generates transitions with differences of the quantum number $\delta M=-2$  within a ladder of a constant $J$, as shown in the insets of Figs.\,\ref{fig5}(d)-\ref{fig5}(f).  In the uncorrelated radiance regime, corresponding to Figs.\,\ref{fig5}(h) and \ref{fig5}(i), almost all atoms are repumped to their excited states due to a strong pumping rate. This is in agreement with the result shown in Fig.\,\ref{fig3}(c).

The collective radiance in steady state  originates from the competition between the collective decay and the repumping of the atoms. In the weak pumping regime, the system makes the balance between the collective decay and the repumping on atoms  by a suppressed emission of the atomic ensemble, leading to steady-state subradiance. In the intermediate pumping regime, the rate of pumping is increased, and the balance between the collective decay and the repumping is realized by an enhanced collective emission of the atoms, which leads to steady-state superradiance. In the strong pumping limit,  the strong repumping enables a large number of atoms to be continuously repumped to their excited states, where the rate of pumping is much larger than that of the two-atom decay. This leads to the generation of uncorrelated radiance of the atomic ensemble.

Next, we investigate the  correlation characteristics of the emitted photons from the atomic ensemble. In Sec.II, we have discussed that the cavity mode can be adiabatically eliminated in the bad-cavity limit (i.e., $\kappa_a\gg\gamma$), and the emission of the cavity photons is thus characterized by the collective emission of the atomic ensemble. Then the correlation functions of the emitted photons can be given by calculating the atomic correlation functions.  Thus, in the case of including only the single-atom decay, the equal-time second-order correlation function\,\cite{Glauber1963}  in the steady-state limit is 
 $g^{(2)}_1(0)=\langle \hat{J}_+\hat{J}_+\hat{J}_-\hat{J}_-\rangle/\langle \hat{J}_+\hat{J}_-\rangle^2$\,\cite{Meiser2010Holland2}.  However, this function is not suitable for the system consisting of two-atom decay, where the cavity field $\hat{a}\propto\hat{J}^2$.  In the case of including only  two-atom decay,  the correlation function can be defined as
\begin{align}\label{eq008}
&g^{(2)}_2(0)=\frac{\langle \hat{J}_+^2 \hat{J}_+^2\hat{J}_-^2\hat{J}_-^2\rangle}{\langle \hat{J}_+^2\hat{J}_-^2\rangle^2}. 
\end{align}
In Figs.\,\ref{fig5}(a)-\ref{fig5}(d), we plot the influence of the repumping on the correlation function. It shows that the nearly coherent emitted photons can be obtained in the superradiance regime when  only single-atom decay is considered, but cannot occur in the case of two-atom decay.  In the latter case, the emitted photons of steady state show bunching in three radiance regimes. The dips of the correlation function correspond to the parameter regime of the maximum superradiance in both cases. In addition,  in the case of two-atom decay, there exists a peak of super bunching in the superradiance regime, and this peak cannot be found in the case of single-atom decay. The dips and peaks correspond to the range of the superradiance regime, thus their positions shift along with the increasing pumping rate.


\section{Discussions and Conclusions} 

Regarding the experimental implementations, a superconducting circuit is an ideal candidate for the implementation of the system with two-atom decay. We consider a circuit QED system consisting of two parametrically coupled  superconducting resonators $\hat{a}$ and $\hat{b}$\,\cite{Chang2020SF,  Vrajitoarea2020HG}, with frequencies $\omega_a$ and $\omega_b$, respectively.  The parametric coupling  between the two resonators with coupling strength $\lambda_{ab}$ can induce the transformation between the single microwave photon in  resonator $\hat{a}$ and the microwave photon pair in resonator $\hat{b}$.   An ensemble of  $N$ identical two-level systems (e.g., qubits and ultracold atoms) with frequency $\omega_{\sigma}$ is coupled to the resonator $\hat{b}$ with the collective coupling strength $\lambda_{b\Gamma}=\sqrt{N}\lambda_{b\sigma}$\,\cite{Kakuyanagi2016MD,Hattermann2017BL,Kubo2011OB, Amsuss2011KN,Kubo2012DD,Putz2014KA,Astner2017NP}, where $\lambda_{b\sigma}$ is the coupling strength between the individual spin and resonator $\hat{b}$, and the frequency of the resonator $\hat{b}$  is far greater than that of the two-level systems.  Under the conditions of $\omega_a\approx 2\omega_{\sigma}$ and  $\omega_{b}\gg \omega_{\sigma}$, the effective resonant transition between two excited two-level systems and the single-photon resonator $\hat{a}$ can be realized, where the transition rate  $\lambda=\lambda_{ab} \lambda_{b\Gamma}^2/[N(\omega_b-\omega_{\sigma})^2]$\,\cite{Qin2021MJ}. Considering the dissipation of the system through coupling to a reservoir, a new two-atom decay channel emerges and the single-atom decay is suppressed.  Here,  the subsequent loss of microwave photons of the system leads to the collective decay of the ensemble of two-system systems. The pumping on the two-level systems can be achieved by optically driving a Raman transition from the ground state $|g\rangle$ to an auxiliary excited state that can rapidly decay to the excited state $|e\rangle$\,\cite{Bohnet2012CW}. The correlation function of emitted microwave photons can be measured by applying quadrature amplitude detectors\,\cite{Bozyigit2011LS, Lang2011BE,Brown1956Twiss}. In this design, we can obtain the correlation of emitted photons $g_2^{(2)}\approx6.22$ with feasible experimental parameters ($N=100$, $\lambda_{ab}/2\pi=\lambda_{b\Gamma}/2\pi=20\,{\rm MHz}$, $(\omega_b-\omega_{\sigma})/2\pi=200\,{\rm MHz}$, $\kappa_a/2\pi=1.6\,{\rm MHz}$, $\kappa_b/2\pi=10\,{\rm KHz}$, $\lambda/2\pi=2\,{\rm KHz}$, $\Gamma_2=4\lambda^2/\kappa_a=10\times 2\pi\,{\rm Hz}$, and $\gamma_p/2\pi=1\,{\rm KHz}$)\,\cite{Vrajitoarea2020HG,Kakuyanagi2016MD,Hattermann2017BL, Kubo2011OB}, where $\kappa_a$ and $\kappa_b$ are the rates of decay of resonators $\hat{a}$ and $\hat{b}$, respectively.   Note that our model is not limited to a particular architecture and could be implemented or adapted in a variety of platforms, such as a waveguide QED system\,\cite{Wang2020JK}. 

In summary, we have investigated the collective radiance characteristics of an atomic ensemble with two-atom decay. The system with two-atom decay relaxes to lower energy states faster than one with single-atom decay.  As a result, the two-atom decay significantly suppresses the steady-state collective radiance of the atomic ensemble in a wide parameter regime, and expands the region of the steady-state subradiance of the atoms.  In particular, compared with the case of  single-atom decay where  the nearly coherent emitted photons can be obtained in the superradiance regime, the emitted photons of the steady state only show bunching in three radiance regimes when only the two-atom decay is considered.   

We thank Dr. C.-S. Hu and Q.-Y. Qiu for helpful discussions. This work is supported by the National Key Research and Development Program of China grant 2021YFA1400700, the National Natural Science Foundation of China  (Grants No.\,11974125, No.\,12205109)  and the China Postdoctoral Science Foundation No. 2021M701323.

\end{document}